\documentclass[10pt,aps,prl,floatfix,twocolumn,nofootinbib,superscriptaddress]{revtex4-2}
\usepackage{mymacros}

\begin{document}
\title{\texorpdfstring{T${}^2$}{} deformations in the double-scaled SYK model: Stretched horizon thermodynamics}
\author{Sergio E. Aguilar-Gutierrez}
\email{sergio.ernesto.aguilar@gmail.com}
\affiliation{Okinawa Institute of Science and Technology Graduate University, Onna, Okinawa 904 0495, Japan}
\affiliation{Institute for Theoretical Physics, KU Leuven, Celestijnenlaan 200D, B-3001 Leuven, Belgium}
\begin{abstract}
We study T$^2$-deformations in the DSSYK model \emph{after} performing ensemble averaging, to learn about the properties of its bulk dual geometry according to finite cutoff holography. We first derive the flow equation for generic holographic quantum mechanical systems dual to dilaton gravity theories without requiring AdS asymptotic boundaries, resulting in a new dictionary entry relating the boundary deformation parameter to the bulk radial boundary cutoff. Applying the results to the DSSYK model, we show that the T$^2$ deformation can probe regions of positive and approximately constant curvature in sine dilaton gravity. The dS$_2$ stretched horizon holographic proposal by Susskind \cite{Susskind:2021esx} is recovered as the radial cutoff approaches the horizons of the bulk dual. The thermodynamic quantities of the system exhibit enhanced growth in this limit. In the deformed boundary theory, we uncover phase transitions from thermodynamically stable to unstable configurations depending on its temperature, which also scrambles information at a (hyper-)fast rate in terms of its out-of-time-ordered correlator.
\end{abstract}
\maketitle

\paragraph{Introduction}\label{sec:intro}
Explicit examples of the holographic duality outside the anti-de Sitter (AdS) space/conformal field theory (CFT) correspondence are relatively scarce and, arguably, not well-understood. Low-dimensional quantum gravity allows new opportunities to explore holography involving bulk geometries with more general metrics and asymptotic boundary conditions from first principles. In particular, the double-scaled SYK model (DSSYK) \cite{Erdos:2014zgc,Cotler:2016fpe,Berkooz:2018jqr,Berkooz:2018qkz} (see \cite{Berkooz:2024lgq} for a recent review) has catalyzed progress in non-AdS holography \cite{Blommaert:2023opb,Blommaert:2024ymv,Almheiri:2024xtw} including different connections to the static patch of dS space (see e.g.~\cite{Verlinde:2024znh,Verlinde:2024zrh,Narovlansky:2023lfz,Susskind:2021esx,Susskind:2022bia,Susskind:2022dfz,Susskind:2023hnj,Rahman:2022jsf,Rahman:2023pgt,Rahman:2024iiu,Rahman:2024vyg,Lin:2022nss,Yuan:2024utc,Aguilar-Gutierrez:2024nau,Xu:2024hoc,Lin:2022rbf}). In this work, we will incorporate T${}^2$ deformations \cite{Hartman:2018tkw}
in the DSSYK model \emph{after} ensemble averaging, 
and contrast its properties with those expected in stretched horizon holography of dS$_2$ space \cite{Susskind:2021esx}. 

\paragraph{The SYK model}
Consider a system with $N$ Majorana fermions in $(0+1)$-dimensions, and $p$-body all-to-all interactions. The theory, developed in several works \cite{kitaevTalks1,kitaevTalks2,kitaevTalks3,Cotler:2016fpe,Maldacena:2016upp,Saad:2018bqo,Maldacena:2018lmt,Jensen:2016pah,Polchinski:2016xgd}, is given by
\begin{equation}\label{eq:DDSSYK Hamiltonian}
    \hat{H}_{\rm SYK}=\rmi^{p/2}\sum_{I}J_{I}\hat{\psi}_{I}~,
\end{equation}
where $I=(i_1,\ldots,i_p)$ is a collective index, with $1\leq i_1<\dots<i_p\leq N$; and similarly $\hat{\psi}_I\equiv\hat{\psi}_{i_1}\dots\hat{\psi}_{i_p}$ is a string of Majorana fermions, where $\qty{\hat{\psi}_{i},~\hat{\psi}_{j}}=2\delta_{ij}$ and the coupling constants $J_I=J_{i_1\dots i_p}$ are Gaussian distributed
\begin{equation}\label{eq:GEA J}
\expval{J_{I}}=0~,\quad\expval{J_{I}J_{J}}=\begin{pmatrix}
        N\\
        p
    \end{pmatrix}^{-1}\frac{\mathcal{J}^2N\delta_{IJ}}{2p^2}~,
\end{equation}
with $\mathcal{J}\in\mathbb{R}$ is an arbitrary constant. Importantly for us, once we consider the double scaling regime, where:
\begin{equation}\label{eq:double scaling}
    N,~ p \rightarrow \infty~,\quad q \equiv \rme^{-\frac{2p^2}{N}}~ \text{fixed}~.
\end{equation}
The SYK model becomes analytically solvable even away from the low-energy regime through an annealed ensemble-averaged description in which the Wick contraction of $\hat{\psi}_I$, represented through chord diagrams \cite{Berkooz:2024lgq}. The Hamiltonian of the ensemble-averaged description of the DSSYK model, $\hH$, has an energy basis $\ket{\theta}$ (where $\theta\in[0,~\pi]$) and eigenvalues given as \cite{Berkooz:2018jqr,Berkooz:2018qkz}
\begin{align}
    &\hH\ket{\theta}=E(\theta)\ket{\theta},\quad E(\theta)=\frac{2\mathcal{J}\cos(\theta)}{\sqrt{\lq(1-q)}}~.\label{eq:energy spectrum}
\end{align}

\paragraph{Stretched Horizon} 
It was conjectured in \cite{Susskind:2021esx} (and further elaborated in \cite{Lin:2022nss,Susskind:2022bia,Rahman:2022jsf,Susskind:2023rxm,Rahman:2024vyg,Susskind:2023hnj,Milekhin:2024vbb}) that the quantum theory dual to dS$_2$ space (seen as a s-wave dimensional reduction from dS$_3$ space) is the (double-scaled) SYK model at infinite temperature residing in a timelike surface very close to the cosmological horizon, called the \emph{stretched horizon}, and that it scrambles information, in terms of the rate of change of out-of-time-ordered correlators (OTOCs) \cite{Rozenbaum:2016mmv}, at a hyper-fast rate (i.e.~faster than exponential; violating the chaos bound \cite{Maldacena:2015waa}). This seems to be in tension with the sine dilaton gravity theory dual to the DSSYK model (at the disk level). However, we will show that there are specific regimes where the 1-dimensional T${}^2$ deformation \cite{Hartman:2018tkw,Gross:2019ach,Gross:2019uxi} of the DSSYK model reproduces the properties conjectured in \cite{Susskind:2021esx}. The T${}^2$ deformation generates a cutoff surface with Dirichlet boundary conditions in the bulk of sine dilaton gravity that can be associated with a stretched horizon.

\begin{figure*}[t!]
	\centering
	\subfloat[]{\includegraphics[height=0.21\textwidth,valign=t]{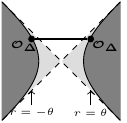}}\hspace{1cm}\subfloat[]{\includegraphics[height=0.21\textwidth,valign=t]{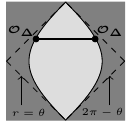}}\hspace{1cm}\subfloat[]{\includegraphics[height=0.21\textwidth,valign=t]{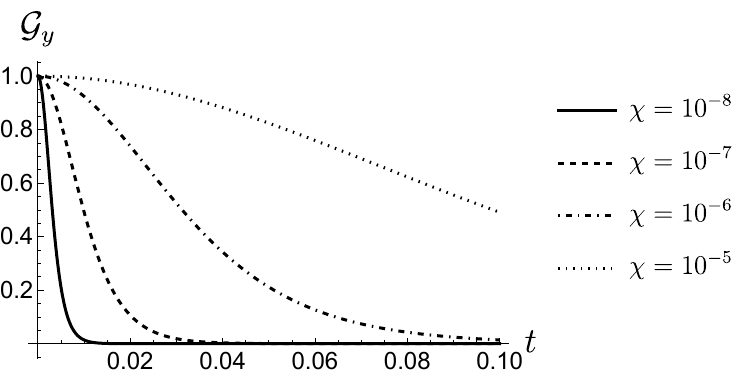}}
	
	\caption{Two-point correlation function of the T${}^2$ deformed DSSYK model (\ref{eq:correlator Lorentzian}). (a, b) Penrose diagrams displaying the insertion of matter operators $\mathcal{O}_\Delta$ at the boundaries close to the horizons at: (a) $r=\pm\theta$, and (b) $r=\theta$ and $r=2\pi-\theta$ (where $\beta_y(\theta)\rightarrow-\beta_y(\theta)$ in (\ref{eq:microcanonical inverse temperature})). The light gray regions in the diagram indicate the positive curvature region between the boundaries, and the part of spacetime that has been cut off is shown in dark gray. (c) Evaluation of (\ref{eq:correlator Lorentzian}) using (\ref{eq:def stretched horizon}, \ref{eq:def parameter}, \ref{eq:microcanonical inverse temperature}) with $\lq=10^{-4}$, $\Delta=1$, $\theta=\frac{\pi}{2}+0.01$, and fixed values of $\chi$. Note an enhanced decay of $\mathcal{G}_y$ when $\chi\ll1$.}
	\label{fig:correlator_time}
\end{figure*}

For this purpose, we study the semiclassical thermodynamics and correlation functions of the DSSYK model placed on the stretched horizon of a positive curvature region in the bulk (see Fig.~\ref{fig:correlator_time} (a, b)). Concretely, we address the question:
\begin{quote}
\emph{By locating the DSSYK model near the horizons in sine dilaton gravity, how do its thermodynamic properties and scrambling of information compare to those expected in the stretched horizon conjecture?}
\end{quote}
We answer this question starting from generic dilaton gravity theories at finite cutoff in the bulk and holographic quantum mechanics in the boundary; we then apply the formalism to the DSSYK model and sine dilaton gravity,  and show an explicit realization of Susskind's conjectures on hyper-fast growth, thus providing a bridge between different holographic proposals of DSSYK in the literature. Since the geometry of sine dilaton gravity is complex-valued, the deformation parameter in the boundary theory can also be complex-valued. Nevertheless, the bulk/boundary systems are well-defined in the specific setting of this work. An extended analysis of the Hamiltonian deformations in this work and their bulk interpretation is presented in our companion paper \cite{Aguilar-Gutierrez:2026ogo}.

\paragraph{General dilaton gravity holograms}\label{ssec:bulk dual}
We seek to derive the flow in the spectrum of holographic quantum mechanical systems dual to dilaton gravity theories of the form
\begin{equation}\label{eq:Dilaton-gravity theory}
\begin{aligned}
    I_{\rm E}=&-\frac{1}{2\kappa^2}\int_{\mathcal{M}}\rmd^2x\sqrt{g}\qty(\Phi\mathcal{R}+U(\Phi))\\
    &-\frac{1}{\kappa^2}\int_{\partial\mathcal{M}}\rmd x\sqrt{h}\qty(\Phi_{B} K-\sqrt{G(\Phi_B)}))~,
\end{aligned}
\end{equation}
where $\mathcal{M}$ is the background manifold, {$\kappa^2$ is a constant (where the weak gravity regime corresponds to $\kappa\rightarrow0$}), $\Phi_{B}$ is the boundary value of the dilaton, $K$ the extrinsic curvature of $\partial\mathcal{M}$, $U(\Phi)$ the dilaton potential, and $G(\Phi_{B})$ is an appropriate counter term selected for the on-shell action to be finite in general dilaton gravity theories with asymptotic boundaries at $\Phi=\Phi_B$ \cite{Grumiller:2007ju}
\begin{equation}\label{eq:counterterm}
    G(\Phi_B)=\int^{\Phi_B}U(\Phi)\rmd\Phi~.
\end{equation}
The Euclidean vacuum solution to the equations of motion (EOM) of (\ref{eq:Dilaton-gravity theory}) for general potential $U(\Phi)$ is:
\begin{equation}\label{eq:metric}
    \rmd s^2=F(r)\rmd\tau^2+\frac{\rmd r^2}{F(r)}~,\quad \Phi=r~,
\end{equation}
where we will allow for $r\in\mathbb{C}$. 
Combining (\ref{eq:metric}) and (\ref{eq:Dilaton-gravity theory}) results in the following on-shell action:
\begin{equation}\label{eq:on shell action}
    I_{\rm E}^{\rm (on)}=-\frac{\pi}{\kappa^2}(\Phi_0+\theta)-\frac{\beta}{2\kappa^2} \qty(F(\Phi_{B})-\sqrt{F(\Phi_{B})~G(\Phi_{B})})~,
\end{equation}
where we have used the fact that $\mathcal{M}$ has a disk topology when $\Phi_0\gg\abs{\Phi}$. The corresponding blackening factor is
\begin{equation}\label{eq:blackening standard}
    F(r)=\int_{\theta}^rV(r')\rmd r'~,
\end{equation}
where $\theta$ denotes a parametrization of the horizon (which in the sine dilaton gravity/DSSYK context corresponds to $\theta$ in \eqref{eq:energy spectrum}).

\paragraph{Flow equation}It was realized in \cite{Gross:2019ach}, based on higher dimensional finite cutoff holography \cite{Hartman:2018tkw}, that moving the radial location of the Dirichlet boundary $r_B$ in generic dilaton gravity theories results in modifications to the energy flow equation of the deformed boundary dual. 
In the Appendix, we generalize the derivation in \cite{Gross:2019ach} to allow asymptotic boundary conditions which are beyond those of AdS$_2$ spacetimes, and also applicable when the dilaton and metric are complex-valued. We find that under radial cutoff flow in dual dilaton gravity of the form \eqref{eq:Dilaton-gravity theory}, the energy spectrum of the deformed theory, $E_y(\theta)$, obeys
\begin{equation}\label{eq:flow eq}
    {\partial_{y} E_y=\frac{E_y^2+(\eta-1)/y^2}{2(1-y E_y)}~,}
\end{equation}
where we have defined the deformation parameters
\begin{equation}\label{eq:def parameter}
    y\equiv2\kappa^2/G(\Phi_{B})~,\quad \eta\equiv U(\Phi_B)/G'(\Phi_B)~.
\end{equation}
The flow equation (\ref{eq:flow eq}) takes the same form as other places in the literature (see e.g.~\cite{Gross:2019uxi}) where $\eta=U(\Phi_B)/G'(\Phi_B)=+1$ corresponds to a 1-dimensional analog of $\text{T}\overline{\text{T}}$ deformations, and similarly $\eta=-1$ to $\text{T}\overline{\text{T}}+\Lambda_2$ deformations \cite{Gorbenko:2018oov,Coleman:2021nor}. We address the general case where $y\in\mathbb{C}$ and $\eta=+1$ in the following, and the $\eta=-1$ case can be found in our companion work \cite{Aguilar-Gutierrez:2026ogo}.

\paragraph{Finite cutoff sine dilaton gravity}We are interested in applying the previous relations to describe the flow in the energy spectrum of the DSSYK model under a T${}^2$ deformation. We adopt sine dilaton gravity \cite{Blommaert:2024ymv,Blommaert:2023wad,Blommaert:2023opb} as the specific dilaton gravity model \eqref{eq:Dilaton-gravity theory} with 
\begin{equation}
    \label{eq:potential}
    U(\Phi)=2\sin\Phi~.
\end{equation}
Recently, it has been realized that there is an isomorphism relating the DSSYK model and the canonically quantized sine dilaton gravity theory at the disk topology level \cite{Blommaert:2023opb,Blommaert:2024ymv,Blommaert:2023wad}. For this reason, when $U(\Phi)=2\sin\Phi$, we will consider the following boundary conditions,\footnote{The on-shell action (\ref{eq:on shell action}) with (\ref{eq:blackening standard}) $F(r)=2\qty(\cos\theta-\cos r)$ in the $\Phi_B\rightarrow\pm(\pi/2+\rmi\infty)$ limit can be used to match the DSSYK energy spectrum (\ref{eq:energy spectrum}) with the Arnowitt–Deser–Misner (ADM) energy \cite{arnowitt1959dynamical} in sine dilaton gravity when:
\begin{equation}\label{eq:match energy bulk bdry}
   {{\kappa^2}={\sqrt{\lq(1-q)}}/{(4\mathcal{J})}~.}
\end{equation}}
\begin{equation}\label{eq:bdry cond}
    \Phi_{B}^{(\pm)}=\pm\qty(\frac{\pi}{2}+\rmi\infty)~,\quad \sqrt{h}~\rme^{\pm \rmi\Phi^{(\pm)}_{B}/2}=\rmi~.
\end{equation}
Both values of $\Phi_B^{(\pm)}$ reproduce the same on-shell observables (see e.g.~(\ref{eq:on shell action})) given that the $\Phi_B$ variation of the boundary action in (\ref{eq:Dilaton-gravity theory}) fixes $K=\frac{U(\Phi_B)}{2\sqrt{G(\Phi_B)}}$, which is invariant under $\Phi_B\rightarrow-\Phi_B$. This leads to an equality between the partition function of the seed boundary theory and the bulk. In finite cutoff holography, the equality extends to the deformed boundary theory and the bulk with a finite radial cutoff coordinate (see the Appendix) from which the flow equation \eqref{eq:flow eq} follows, and the transfer matrix of the deformed boundary theory becomes 
\begin{equation}
H_y=\frac{1}{y}\qty(1-\sqrt{1-2y \hH})~.
\end{equation}
Note that the holographic dictionary \eqref{eq:def parameter} is preserved under $\Phi_B\rightarrow-\Phi_B$ when $U(\Phi)=2\sin\Phi$, which is a symmetry in the action. We will then use the \emph{deformation parameter} $y$ to locate the deformed DSSYK model at $r_{B}$.
  
We exploit that the background (\ref{eq:metric}) interpolates between negative and positive curvature regions since $\mathcal{R}=-2\cos r$. 
In what follows, $\theta$ is a fixed bulk parameter, and the dual deformed DSSYK model is described by fixed energy state $\ket{E(\theta)}$.

As anticipated above, the bulk geometry near the horizons at $r=2\pi\pm\theta$ approximates dS$_2$ space when $\theta\simeq\pi$ since $\mathcal{R}\simeq 2$, and it indeed possesses similar properties to the conjecture in \cite{Susskind:2021esx}, including the (hyper-)fast scrambling of information (described below (\ref{eq:MilenkinXu})). 
We will probe the thermodynamic and scrambling properties of the positive curvature regions near the horizons using $r=r_B(\theta)$ as the location of (\ref{eq:bdry cond}) given by:
\begin{align}\label{eq:def stretched horizon}
    r_{B}(\theta)=(1-\chi)\theta+\chi(\theta-2\pi)~.
\end{align}
Here $\chi\in[0,~1]$ is a parametrization that allows us to interpolate the location between consecutive horizons (i.e.~where $F(\theta)=0$). This parametrization has been used extensively in static patch stretched horizon holography \cite{Susskind:2021esx,Jorstad:2022mls,Baiguera:2023tpt,Aguilar-Gutierrez:2024rka,Baiguera:2024xju}; however, the precise parametric form of $r_B(\theta)$ is not relevant, as we are mostly concerned in the near horizon region to probe a nearly-dS$_2$ space. 

\paragraph{Reality conditions}So far, we have allowed the deformation parameter to be complex-valued due to the geometry of sine dilaton gravity being complex-valued. While this generically results in a complex energy spectrum, it does not result in non-unitary Lorentzian time evolution. This can be seen from the $\eta=+1$ spectrum
\begin{equation}\label{eq: E lambda TTbar}
    E_y(\theta)=\frac{1}{y}\qty(1\pm\sqrt{1-2y E(\theta)})~.
\end{equation}
The solution with a relative $-$ sign is smoothly connected to the seed theory in the limit $y\rightarrow0$; while the relative $+$ is a non-perturbative solution. To study the thermodynamic properties of the deformed DSSYK model, we will focus on the solution with relative $-$ sign in (\ref{eq: E lambda TTbar}), while we consider both $\pm$ in the analysis of correlation functions, as they contain complementary information to describe the positive curvature regions in sine dilaton gravity.

\begin{figure*}
    \centering
   \subfloat[]{\includegraphics[height=0.25\textwidth]{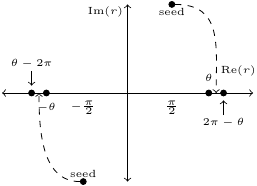}}\hspace{1cm}\subfloat[]{\includegraphics[height=0.25\textwidth]{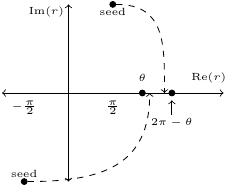}}
    \caption{The DSSYK model is initially located at $r\rightarrow\pm(\frac{\pi}{2}+\rmi\infty)$, and the T${}^2$ deformation allows the corresponding boundary location to move along the indicated contours (dashed lines) close to the horizons at (a) $r=\pm\theta$, and (b) $r=\theta,~2\pi-\theta$. Both ways can be used to probe the positive curvature regions between the horizons when $\theta\simeq\pi$ (see Fig.~\ref{fig:correlator_time}).}
    \label{fig:Contour_TTbar}
\end{figure*}

Consider the radial bulk flow illustrated in Fig.~\ref{fig:Contour_TTbar}. First, note that as we flow the boundary theory from  $r_B\rightarrow\pm\qty(\frac{\pi}{2}+\rmi\infty)$ to generic locations $r_B\in\mathbb{C}$ in Fig.~\ref{fig:Contour_TTbar}, the energy spectrum (\ref{eq: E lambda TTbar}) becomes complex;\footnote{The deformation parameter can also become complex in traditional AdS black hole spacetimes. This happens for instance in Cauchy slice holography \cite{Araujo-Regado:2022gvw}, where the boundary stress tensor is allowed to be complex-valued, resulting in a complex energy spectrum. However, similar to our setting, the non-unitary evolution is avoided due to the metric being Euclidean.} however, the metric (\ref{eq:metric}, \ref{eq:blackening standard}) is complex-valued along this flow, and there is no Lorentzian description where we could associate the complex energy spectrum to, e.g.~a dissipation process. Therefore, once $r_B$ becomes real, as in Fig.~\ref{fig:Contour_TTbar}, the model evolves unitarily as long as $y E(\theta)\leq 1/2$.

Next, consider the deformed DSSYK partition function
\begin{equation}\label{eq:partition DSSYK}
	Z_y(\beta)\equiv\expval{\tr(\rme^{-\beta H_{\rm SYK}})}_y=\bra{0}\rme^{-\beta \hH_y}\ket{0}~,
\end{equation}
where $\ket{0}$ indicates the empty open chord state in the ensemble-averaged theory. We note that even when the exponent of the integral is complex-valued, given that the deformed energy spectrum is bounded, the partition function remains bounded, so that the theory remains well-behaved when $y$ \eqref{eq:def parameter} is complex-valued.

\paragraph{Semiclassical thermodynamics}
We will now study the semiclassical limit ($\lq\ll1$) of \eqref{eq:partition DSSYK} including one-loop corrections in the thermodynamic entropy studied in \cite{Goel:2023svz}. Following the evaluation in \cite{Goel:2023svz} (\ref{eq:partition DSSYK}) becomes
\begin{equation}\label{eq:partion function a}
	Z_y(\beta)=\int\rmd E(\theta)~\rme^{S(\theta)-\beta E_y(\theta)}~,
\end{equation}
where the thermodynamic entropy is
\begin{equation}\label{eq:thermodynamic entropy}
	S(\theta)=\frac{1}{\lq}\qty(2\pi\Phi_0+2\theta(\pi-\theta))~,
\end{equation}
and we have introduced $\Phi_0$ from an overall normalization of the partition function. {We can then match the thermodynamic entropy derived from the bulk on-shell action (\ref{eq:on shell action}) when $8\pi G_N=1$,\footnote{To match the $\theta^2$ term in (\ref{eq:thermodynamic entropy}) one needs to introduce bulk constraints dual to a positive chord number condition \cite{Blommaert:2024ymv}.} which we will fix from now on.}
Using (\ref{eq:match energy bulk bdry}) and \eqref{eq:thermodynamic entropy}, we identify the inverse temperature in the microcanonical ensemble:
\begin{equation}\label{eq:microcanonical inverse temperature}
	\beta_y(\theta)\equiv\dv{S}{E_y}=\beta(\theta)\sqrt{1-2y E(\theta)}~, \quad \beta(\theta)\equiv \frac{2\pi-4\theta}{\sin\theta}~.
\end{equation}
It can then be verified that the energy (\ref{eq: E lambda TTbar}) and thermodynamic entropy (\ref{eq:thermodynamic entropy}) experience an enhanced growth as ${\beta_y\rightarrow0}$ for $\chi\ll1$, while it remains bounded in the range where $E_y\in\mathbb{R}$.

Moreover, we identify a dictionary between the boundary and bulk thermodynamic parameters by equating the on-shell partition function on both sides using the bulk action (\ref{eq:on shell action}) and the saddle-point solution to the boundary free energy, i.e.~$S(\theta)-\beta_y E_y(\theta)$ using (\ref{eq:flow eq}, \ref{eq:match energy bulk bdry}, \ref{eq:microcanonical inverse temperature}). Matching the $\beta$-dependent and $\beta$-independent terms above, we recover:
    \begin{align}\label{eq:beta_T E_rB}
    \beta_T &=\beta\sqrt{F(r_{B})}=\beta_y\frac{\qty(\lq(1-q))^{1/4}}{\sqrt{2\mathcal{J}y}}~,\\
    E_{\rm BY}&=\frac{1}{2\kappa^2}\qty(\sqrt{G(r_{B})}-\sqrt{F(r_{B})})=\frac{\sqrt{2\mathcal{J}y\, E_y}}{\qty(\lq(1-q))^{1/4}}~.\nonumber
\end{align}
where the first line is the inverse \emph{Tolman temperature} and the second is the \emph{Brown-York energy} \cite{Brown:1992br}. These parameters are in general redshifted (by the factor $F(r_B)$ in \eqref{eq:beta_T E_rB}) with respect to the asymptotic boundary limit. One can appreciate that \eqref{eq:beta_T E_rB} identifies the local thermodynamic parameters that an observer in a gravitating spacetime detects in terms of purely boundary data. This information could be used e.g.~as a consistency check in the derivation of the thermodynamic first law on both sides of the finite cutoff holographic correspondence \cite{Zhang:2025dgm}.

\begin{figure*}[t!]
    \centering
 \includegraphics[height=0.2\textwidth]{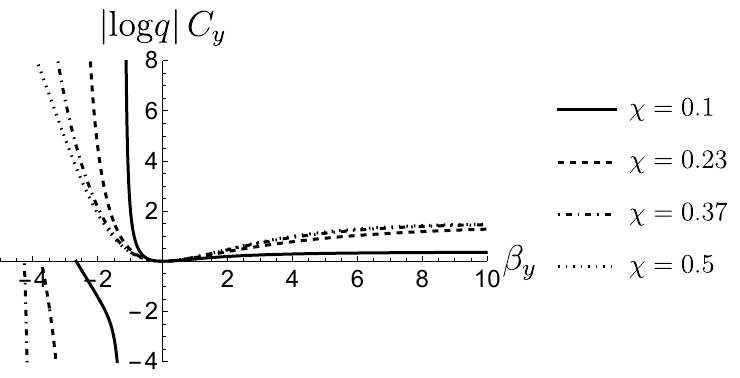}\hspace{0.5cm}\includegraphics[height=0.2\textwidth]{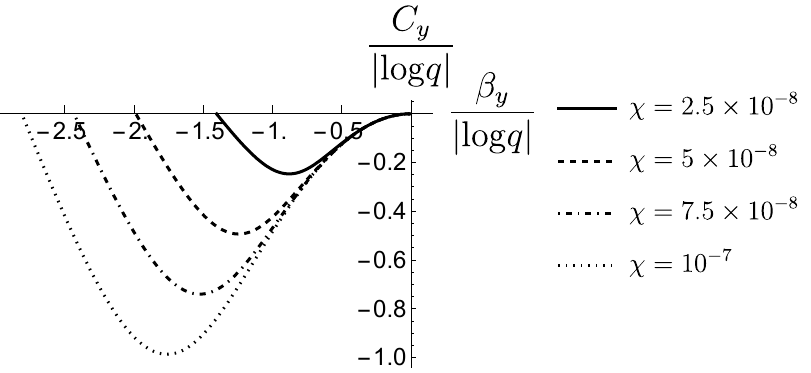}

    \caption{Heat capacity (\ref{eq:heat capacity}) of the deformed DSSYK model as a function of the temperature after rescaling by a factor of $\lq$, indicated in the corresponding axis, for display purposes. We have taken $\lq=10^{-3}$ and different values of the $\chi$ parameter defined in (\ref{eq:def stretched horizon}). The results display a transition between thermodynamically stable and unstable phases.}
    \label{fig:energy_entropy}
\end{figure*}

\paragraph{Phase transition}
Next, we determine the thermal stability of the deformed theory, using the heat capacity
\begin{equation}\label{eq:heat capacity}
		C_{y}\equiv-\beta_y^2\dv{E_{y}}{\beta_{y}}~,
  \end{equation}
  where we specialize in sine dilaton gravity (\ref{eq:potential}) with a stretched horizon (\ref{eq:def stretched horizon}),  resulting in the plots in Fig.~\ref{fig:energy_entropy} as a function of $\beta_y$ (\ref{eq:microcanonical inverse temperature}).
We have fixed the relative distance between the Dirichlet wall and the horizon location as we vary the microcanonical temperature (\ref{eq:microcanonical inverse temperature}). From the dictionary (\ref{eq:beta_T E_rB}) we also identify $C_y$ as the heat capacity at fixed $r_B$ in the bulk \cite{Aguilar-Gutierrez:2026ogo}. We observe that for generic values of $\chi$, $C_{y}$ changes sign at specific values of $\beta_y$, which indicates that the system undergoes a phase transition from a thermodynamically stable to an unstable configuration due to the change of sign in $C_{y}$. In contrast, when the boundary theory is very close to the cosmological horizon, the system is always thermodynamically unstable. This is in favor of Susskind's conjecture \cite{Susskind:2021esx}; as we probe the nearly dS$_2$ region in sine dilaton gravity, and zoom in on the corresponding horizon, the system displays the thermodynamic instability of the static patch of dS$_2$ space with Dirichlet boundaries.

\paragraph{Thermal two-point correlation functions}
We consider adding matter chord operator insertions in the path integral to compute the (un-normalized) Euclidean two-sided thermal two-point correlation function as
\begin{align}\label{eq:path int correlator TTbar}
    &G_y(\tau)\equiv\expval{\tr({\rme^{-(\beta-\tau) \hH_{{\rm SYK}}}\mathcal{O}_\Delta\rme^{-\tau \hH_{{\rm SYK}}}\mathcal{O}_\Delta})}_{y}~,\\
    &\mathcal{O}_\Delta\equiv\rmi^{\frac{p'}{2}}\sum_{I'}K_{I'}\psi_{I'},\quad \expval{K_{I'}K_{J'}}=\binom{N}{p'}^{-1}\frac{\mathcal{K}^2\delta_{I'J'}}{\Delta^2\lq}~,\nonumber
\end{align}
where $I'=i_1,\dots,i_{p'}$, $\Delta\equiv p'/p$; the trace in the first line is taken in the microscopic Hilbert space of the Majorana fermions $\psi_i$; and $K_{I'}$ are Gaussian random couplings that are independent of $J_{I}$; $\mathcal{K}$ is a constant (normalized as $\expval{\tr(\mathcal{O}_{\Delta}^2)}=1$ from now on), while the T${}^2$ deformation is performed after taking the ensemble-averaging in (\ref{eq:path int correlator TTbar}). Evaluating (\ref{eq:path int correlator TTbar}) at the saddle point leads to \cite{Aguilar-Gutierrez:2026ogo}
\begin{equation}\label{eq:correlator Euclidean}
	\lim_{\lq\rightarrow0}G_y(\tau)=\qty(\frac{\sin\theta}{\sin(\theta+\frac{\pi-2\theta}{\beta_y(\theta)}\tau)})^{2\Delta}~,
\end{equation}
where $\tau\leq\abs{\beta_y(\theta)}$. Note that the time-dependence of the correlator is the same as the $y=0$ case upon $\beta\rightarrow\beta_y$, which is consistent with the dictionary relating boundary and bulk thermodynamics in (\ref{eq:beta_T E_rB}).

\paragraph{Tomperature} The Lorentzian continuation $\tau\xrightarrow{}\rmi t+\frac{\beta_y(\theta)}{2}$ in (\ref{eq:correlator Euclidean}) results in the thermal correlation function, which we normalized as $\mathcal{G}_y(t=0)=1$:
\begin{equation}\label{eq:correlator Lorentzian}
	\mathcal{G}_y(t)={\sech^{2\Delta}\qty(\frac{\pi-2\theta}{\beta_y(\theta)}t)}~.
\end{equation}
The decay rate of the thermal correlation function (sometimes called ``tomperature'' \cite{Lin:2022nss}, or fake temperature \cite{Lin:2023trc}) can be read immediately from (\ref{eq:correlator Lorentzian}) as $\abs{(\pi-2\theta)/\beta_y}$.

\paragraph{Bulk interpretation}
From the bulk perspective, the positive curvature regions are highlighted in Fig.~\ref{fig:correlator_time} (a, b), where we interpret the addition of the matter chord operators in terms of bulk scalar operators with conformal weight $\Delta$ in sine dilaton gravity \cite{Blommaert:2024ymv}.
Here, Fig.~\ref{fig:correlator_time} (a) corresponds to the deformed DSSYK model with the energy spectrum in (\ref{eq: E lambda TTbar}) with a relative $-$ sign, which gives us access to positive curvature region $r\sim\pm\pi$. In contrast, Fig.~\ref{fig:correlator_time} (b) corresponds to the deformed DSSYK model with a relative $+$ sign in (\ref{eq: E lambda TTbar}) (resulting in a change of sign $\beta_y(\theta)\rightarrow-\beta_y(\theta)$ in (\ref{eq:microcanonical inverse temperature})) where the boundary location is given in (\ref{eq:def stretched horizon}). This allows us to probe the positive curvature region that has no horizons in between. Lastly, Fig.~\ref{fig:correlator_time} (c) illustrates the Lorentzian evolution of the correlation function for both (a) and (b), given that (\ref{eq:correlator Lorentzian}) is invariant under $\beta_y(\theta)\rightarrow-\beta_y(\theta)$. The autocorrelation function is not sensitive to the negative curvature region between the horizons in Fig.~\ref{fig:correlator_time} (a) with respect to (b).

\paragraph{Hyper-fast scrambling}\label{ssec:hyperfast}
As displayed in Fig.~\ref{fig:correlator_time} (c), once we take the radial boundary cutoff to be given by (\ref{eq:def stretched horizon}), the normalized correlator in the $\chi\rightarrow0$ limit (as well as for $\chi\rightarrow1$) decays essentially immediately. In fact, it has been shown in \cite{Milekhin:2024vbb} that the OTOCs (defined by the left-hand side of (\ref{eq:MilenkinXu})) involving a general pair of products of fermionic operators at infinite temperature is bounded from below by the corresponding two-point function of the same operators. In terms of the chord matter operators in the Heisenberg picture $\mathcal{O}_\Delta(t)=\rme^{\rmi \hH t}\mathcal{O}_\Delta\rme^{-\rmi \hH t}$, the statement translates into:
\begin{equation}\label{eq:MilenkinXu}
    \abs{\expval{\tr(\mathcal{O}_\Delta(t)\mathcal{O}_\Delta(0)\mathcal{O}_\Delta(t)\mathcal{O}_\Delta(0))}}\geq 2\mathcal{G}_y(t)^2-1~.
\end{equation}
Fig.~\ref{fig:correlator_time} (c) together with (\ref{eq:MilenkinXu}) show that the scrambling of the deformed DSSYK model at the stretched horizon, as measured by the decay rate of the OTOC in the infinite temperature limit, exhibits an enhanced exponential decay rate, confirming the hyper-fast scrambling property of the quantum theory living on the stretched horizon conjectured in \cite{Susskind:2021esx}, even though the background is not dS$_2$ space. We present a more detailed OTOC analysis in \cite{Aguilar-Gutierrez:2026ogo}, including the growth near the scrambling time.

\paragraph{Discussion} We studied 1-dimensional Hamiltonian deformations derived from finite cutoff holography in generic dilaton gravity theories. We applied the general formulation to probe the positive curvature regions in the sine dilaton gravity from T${}^2$ deformations of the DSSYK model, which we related to the stretched horizon holographic proposal in \cite{Susskind:2021esx}, thus connecting two different bulk dual proposals of the DSSYK model. While the spacetime geometry of sine dilaton gravity is generically different from the static patch of dS$_2$ space, one can use a radial bulk flow (Fig.~\ref{fig:Contour_TTbar}) to isolate specific regions with an approximately constant and positive scalar curvature.\footnote{A different approach in the literature is to describe the Milne patch of dS$_2$ by zooming onto the high energy edge of the spectrum \cite{Blommaert:2023opb,Blommaert:2024whf,Heller:2025ddj,Aguilar-Gutierrez:2025otq,Aguilar-Gutierrez:2026ogo}.} Even though the deformed energy spectrum generically becomes complex-valued during the T${}^2$ flow, the bulk spacetime geometry is also complex (i.e.~there is no Lorentzian evolution) until one isolates the region described by unitary evolution, and the bulk/boundary partition functions remain well-defined due to UV finiteness of this specific setting. By placing the T${}^2$ deformed DSSYK model on the Dirichlet boundaries near the horizons in sine dilaton gravity, we evaluated the thermodynamic entropy, the energy spectrum, heat capacity, and the decay rate of correlators. In particular, the deformed DSSYK model displays a phase transition depending on its separation with respect to the bulk horizons and its temperature, from a thermodynamically stable configuration to an unstable one. Meanwhile, when the cutoff approaches the horizons, the theory is always unstable. Moreover, the above thermodynamic observables display a non-trivial enhanced rate of growth in the near-infinite temperature regime. Our results, combined with a general bound on OTOCs in \cite{Milekhin:2024vbb}, show there is hyper-fast scrambling expected by \cite{Susskind:2021esx}. In \cite{Aguilar-Gutierrez:2026ogo}, we elaborate on and extend these results.

\paragraph{Outlook}There are several future research avenues. One of the most pressing ones is to formulate a microscopic (i.e.~without ensemble averaging) interpretation of our results. 
It is particularly interesting to investigate the thermodynamic instability of the deformed DSSYK within the microscopic setting. We also aim to explicitly relate our results to new ADM Hamiltonians in JT gravity at finite cutoff \cite{Griguolo:2025kpi,Griguolo:2026eoi} which might be most transparent when using a chord Krylov basis \cite{Aguilar-Gutierrez:2026ogo}.

\begin{acknowledgments}
\noindent \emph{Acknowledgments:} I thank Andrew Svesko, and Manus Visser for collaboration on a related work \cite{Aguilar-Gutierrez:2024nst}; Takanori Anegawa, Andreas Blommaert, Klaas Parmentier, and Jiuci Xu for important correspondence; as well as Dio Anninos, Luis Apolo, Arghya Chattopadhyay, Victor Franken, Damián A. Galante, Monica Guica, Eleanor Harris, Andrew Rolph, and Gopal Yadav for useful conversations. SEAG was partially supported by the FWO Research Project G0H9318N and the inter-university project iBOF/21/084, and is supported by the Okinawa Institute of Science and Technology Graduate University. This project/publication is also made possible through the support of the ID\#62312 grant from the John Templeton Foundation, as part of the ‘The Quantum Information Structure of Spacetime’ Project (QISS), as well as Grant ID\# 62423 from the John Templeton Foundation. The opinions expressed in this project/publication are those of the author(s) and do not necessarily reflect the views of the John Templeton Foundation.
\end{acknowledgments}

\appendix

\section{Notation}
We summarize the main notation in alphabetical order in the second column.
\begin{itemize}[noitemsep,
    leftmargin=1.7cm,
    labelwidth=1.6cm,
    labelsep=0.1cm,
    align=left]
\item[$F$] Blackening factor.
\item[$G$] Boundary counterterm.
\item[$T_{ab}$] Boundary matter stress tensor.
\item[$\gamma_{ab}$] Boundary metric.
\item[$\tilde{T}_{ab}$] Bulk matter stress tensor.
\item[$\eta$] Deformation parameter which modifies the bulk geometry, analogous to T$\overline{\text{T}}+\Lambda_2$ deformations, in 1 dimension.
\item[$Z_y$] Deformed partition function.
\item[$U(\Phi)$] Dilaton potential.
\item[$\mathcal{O}_{\Phi_B}$] Dilaton source.
\item[$G_y$] Euclidean time correlator.
\item[$\mathcal{H}$] Hamiltonian constraint.
\item[$h_{\mu\nu}$] Induced metric.
\item[$\mathcal{G}_y$] Lorentzian time correlator.
\item[$\mathcal{O}_\Delta$] Matter chord operator.
\item[$K$] Mean curvature.
\item[$\theta$] Parametrization of the energy spectrum of the DSSYK model.
\item[$q$] $q$-deformation parameter.
\item[$\chi$] Stretched horizon parameter.
\item[$y$] T${}^2$ deformation parameter.
\item[$E_y(\theta)$] T${}^2$ deformed energy spectrum.
\item[$\hH_y$] T${}^2$ deformed Hamiltonian.
\item[$S(\theta)$] Thermodynamic entropy.
\end{itemize}
\section{Deriving the flow equation \texorpdfstring{\eqref{eq:flow eq}}{}}\label{sec:flow eq}
The energy spectrum of a T${}^2$ deformed quantum theory is determined from the location of the radial cutoff boundary in the bulk through the holographic dictionary equating the partition function between both sides of the duality \cite{Hartman:2018tkw}, namely
\begin{equation}
    Z_{\rm bdry}[y,\gamma_{ab},J]=Z_{\rm grav}[r_B,h_{ab},\psi_\Delta]~,
\end{equation}
where $Z_{\rm bdry}$ and $Z_{\rm grav}$ denote the on-shell boundary and bulk theory partition functions respectively; $\gamma_{ab}$ represents the boundary theory metric, $y$ the deformation parameter, $J$ the source for the boundary operator $\mathcal{O}_\Delta$ with conformal dimension $\Delta$; $h_{ab}$ the induced metric at the boundary at the cutoff surface $r_B$, and $\psi_\Delta$ the bulk field of dimension $\Delta$.

The bulk stress tensor used in the evaluation of $Z_{\rm grav}$ is related to the location of the cutoff through the Hamilton-Jacobi (HJ) equation
\begin{equation}\label{eq:HJ eq}
        \partial_{\Phi_{B}}I_{\rm E}^{\rm(on)}+\mathcal{H}\qty({\Phi_{B}};~\frac{\delta I_E^{\rm(on)}}{\delta{\Phi_{B}}})=0~.
    \end{equation}
Here the variation of the on-shell action corresponding to (\ref{eq:Dilaton-gravity theory}) gives
\begin{align}
\label{eq:derivative I E}
        &\partial_{\Phi_{B}}I_{\rm E}^{\rm (on)}=\int_{\partial\mathcal{M}}\rmd\tau\sqrt{h}\qty[\frac{1}{2}\tilde{T}^{ab}\partial_{\Phi_{B}}h_{ab}+\mathcal{O}_{\Phi_{B}}]~,
\end{align}
where
\begin{align}      
&\tilde{T}^{ab}\equiv\frac{2}{\sqrt{h}}\frac{\delta I_E}{\delta h_{ab}}=\frac{2}{\sqrt{h}}\qty(\pi^{ab}+\frac{\sqrt{h}}{4\kappa^2}\sqrt{G(\Phi_{B})}h^{ab})~,\label{eq:BY stress tensor}\\
        &\mathcal{O}_{\Phi_{B}}\equiv\frac{1}{\sqrt{h}}\frac{\delta I_E}{\delta {\Phi_{B}}}=\frac{1}{\sqrt{h}}\qty(\pi_{\Phi_{B}}+\frac{\sqrt{h}~G'({\Phi_{B}})}{4\kappa^2\sqrt{G({\Phi_{B}})}})~,\label{eq:source Phi}
        \\&\pi^{ab}\equiv-\frac{\sqrt{h}}{4\kappa^2}h^{ab}n^c\nabla_c{\Phi_{B}}~,\quad\pi_{\Phi_{B}}\equiv-\frac{\sqrt{h}}{2\kappa^2}K~;
    \end{align}
with $G'({\Phi_{B}})\equiv\dv{G({\Phi_{B}})}{\Phi_{B}}$, and $\mathcal{H}$ in (\ref{eq:HJ eq}) is the Hamiltonian constraint for (\ref{eq:Dilaton-gravity theory}) given by \cite{Grumiller:2007ju}
    \begin{equation}
    \begin{aligned}
        \mathcal{H}&=\frac{1}{2\kappa^2}K~n^c\nabla_c\Phi_B-\frac{1}{4\kappa^2}U(\Phi_{B})=0~.
    \end{aligned}
    \end{equation}
    Similar to \cite{Hartman:2018tkw,Gross:2019ach}, we recognize that to match the boundary metric $\gamma$ with the induced metric at $r_{B}$, their relative scaling, and similarly for the stress tensor takes the form
\begin{equation}
    \sqrt{h}=\sqrt{G(\Phi_{B})}\sqrt{\gamma}~,\quad \tilde{T}_{\tau\tau}=\sqrt{G(\Phi_{B})}~T_{\tau\tau}~.
\end{equation}
Since in (0+1)-dimensions, $T^\tau_\tau=E_y$, which denotes the energy spectrum of the deformed theory, we recover the energy flow equation for the deformed DSSYK model in \eqref{eq:flow eq}
\begin{equation}
    {\partial_{y} E_y=\frac{E_y^2+(\eta-1)/y^2}{2(1-y E_y)}~,}
\end{equation}
where we have defined the deformation parameters in terms of those in the bulk at finite cutoff $y\equiv2\kappa^2/G(\Phi_{B})$ and $\eta\equiv U(\Phi_B)/G'(\Phi_B)$ as in \eqref{eq:def parameter}. Additional details about this derivation are presented in the companion manuscript \cite{Aguilar-Gutierrez:2026ogo}.

\bibliographystyle{apsrev4-1}
\bibliography{references.bib}

\end{document}